\documentstyle[11pt]{article}
\begin{document}
\begin{flushright}
Liverpool Preprint: LTH 298\\
January 27, 1992
\end{flushright}
\vspace{5mm}
\begin{center}
{\LARGE\bf
Lattice study of sphaleron transitions
in a 2D O(3) sigma model}\\[1cm]

{\bf J. Kripfganz}$^1$ and {\bf C. Michael}$^2$\\
\it{$^1$ Fakult\"at f\"ur Physik, Universit\"at Bielefeld,Germany} \\
\it{$^2$ DAMTP, University of Liverpool, Liverpool, L69 3BX, U.K.}

\end{center}

\begin{abstract}
A lattice approach is developed to measure the sphaleron
free energy. Its feasibility is demonstrated through a
Monte Carlo study of the two-dimensional O(3) sigma model.
\end{abstract}
\par
In the electroweak standard model,
baryon number (or, more precisely, $B + L$)
is violated by an anomaly~\cite{thooft}.
Baryon number violation is associated with transitions between
different, topologically distinct vacua.
These vacua are separated by an energy barrier. Sphalerons
\cite{klinkhamer} are  classical solutions corresponding to saddle
points on top of the barrier. The lowest-energy sphaleron has an
energy of the order of 10 TeV (somewhat depending on the
Higgs mass). This energy is concentrated in a region of
size $m_W^{-3}$.
\par
The tunneling rate between
nonequivalent vacua is tiny, due to the small electroweak gauge
coupling. Therefore, baryon number violation in the standard model
was not considered to be of practical importance for some time.
Later, however,
it was realized that the energy barrier
could be overcome more easily by classical transitions
instead of tunneling.
Classical transitions become relevant at
high temperature~\cite{kuzmin},
or perhaps in high energy scattering \cite{ringwald,espinoza}.
\par
The case of
high energy scattering is still badly understood
(see {\it e.g.} Ref. \cite{mattis} for a recent review).
The case of high temperature is extremely interesting because
anomalous baryon number
violation could provide a scenario
for the generation of the baryon number of the universe.
\par
In the high temperature phase with restored symmetry
(B+L) transitions
are believed to be frequent, with a rate
\cite{arnold1,khlebnikov,ambjorn,bochkarev1}
\begin{equation}
\Gamma = \gamma (\alpha_w T)^4
\label{eq:one}
\end{equation}
per unit volume, where
$\gamma$ is some constant of order one.
This would have the striking consequence that any (B+L) asymmetry
generated at a scale close to the Planck mass would be
washed out completely. Where would the baryon asymmetry
of the universe come from in that case?
\par
One  very attractive possibility would
be the generation of a (B+L) asymmetry at the electroweak phase
transition
\cite{shaposhnikov1,bochkarev2,mclerran1,cohen,turok,mclerran2}.
This would  be possible if the electroweak
phase transition would be strongly first order. Presumably, this
requires a more complicated Higgs sector (not just one
doublet), which would be an interesting prediction. The
electroweak phase transition is intensively studied by
perturbative \cite{carrington,dine} as well as lattice
techniques \cite{bunk1,bunk2}.
\par
The survival of such a (B+L) asymmetry after the electroweak
phase transition
is a non-trivial matter, however. It could be washed out
afterwards by
transitions across the  sphaleron.
This will not be the case if the sphaleron barrier near the
critical temperature is high enough.
The corresponding condition on the sphaleron free energy
$F_{sp}(T)$ is found by comparing the transition rate with
the expansion rate of the universe.
\begin{equation}
{{F_{sp} (T_c)}\over T_c} \geq 45,
\label{eq:two}
\end{equation}
where $T_c$ is, strictly speaking, not $T_{crit}$, but the temperature
at which the
phase transition is completed.
In the case of
a second order or weakly first order phase transition this relation will
not be satisfied.
\par
The trouble is that the temperature dependence of the
sphaleron free energy is not very well known, in particular
close to the critical temperature. Perturbation theory is
plagued with severe infrared divergences in this regime. This
should also apply
to the treatment of fluctuations around the
sphaleron. The usual way of estimating the sphaleron free
energy is by rescaling, i.e. replacing the zero-temperature
Higgs expectation value by $ v(T)$, obtained from the
temperature dependent effective potential. This is correct
to leading order but could be quite misleading
because of infrared singularities of higher loop
contributions \cite{kripfganz1}.
The role of the effective potential
itself is also obscure because the sphaleron receives its energy
mostly from the non-convex region which is unphysical.
\par
Therefore a direct determination of
the sphaleron free energy, i.e. the transition rate,  is an important
task. The sphaleron free energy is directly related
to the effective potential of the Chern-Simons number.
A fraction of configurations with CS number close to $1 \over{2}$
will contribute to the anomalous processes in question. The basic
objective is therefore to measure the probability of configurations
close to the sphaleron, and derive the sphaleron free energy.
\par
Determining the probability distribution of the CS number requires
a reliable way of measuring it. In a typical Monte Carlo configuration,
quantum fluctuations (completely unrelated to topological
features) may easily add up to produce a CS number of, e.g., $0.5$.
A straightforward measurement of the CS distribution would
therefore overestimate the rate of fermion number violation.
By focussing on the known properties of the sphaleron,  it is
possible to select configurations which look like an underlying
 classical sphaleron with added quantum fluctuations. Even so it
is necessary to check the stability of such assignments.
 This situation is well
known from lattice estimates of the topological susceptibility
of pure Yang-Mills theory. In that case, cooling techniques
\cite{teper,emi}
have been shown to be useful. Instantons show up as long-lived
states in this way. This method cannot be taken over without
modification to measure the CS density, however, because
sphalerons are not stable classical solutions like instantons,
but saddle points. They cannot be found by simply
minimizing the action. A different smoothing procedure is
needed. We propose to use the square of the tadpole (i.e.
the square of the equation of motion) as a new `action' density of the
cooling procedure. Any classical solution, stable or not, will
now appear as an attraction point~\cite{saddle}.
\par
In order to test such an algorithm one should use a simple
model, where the
magnitude of topological transitions is sufficiently understood.
The Abelian Higgs model in 1+1 dimensions would be a possible
candidate. We prefer the study of the 1+1 dimensional O(3)
sigma model with some external magnetic field, because of its
similarities with the 4-dimensional standard model. The action is
\begin{equation}
S =
\frac{1}{g^2} \int d^2 x \left[ \frac{1}{2} \partial_\mu \vec n \cdot
\partial_\mu \vec n + \omega^2 (1 + n_3) \right]
\label{eq:three}
\end{equation}
with the constraint
\begin{equation}
\vec n^{~2} (x) = 1
\label{eq:threea}
\end{equation}
Without the
external field $\omega$, the theory is asymptotically free and
possesses instanton
solutions. Symmetry breaking due to the external field removes
the instantons as true solutions, but sphalerons as unstable
saddle point solutions appear, just like in the SU(2) Higgs
theory. Topological transitions in this model have been studied
in quite some detail by Mottola and Wipf \cite{mottola}. The one-loop
expression for the transition rate has been worked out and
is expected to be reliable at weak coupling (this is not obvious
in the standard model because of infrared problems near the
phase transition).
To exponential accuracy the transition rate is given by
\begin{equation}
\Gamma \simeq e^{- \frac{F_{sp} (T)}{T}}
\label{eq:four}
\end{equation}
where the sphaleron free energy is found to be
\begin{equation}
F_{SP} (T) = \frac{8\omega}{g^2_R (T)}
\label{eq:five}
\end{equation}
To one-loop order, the renormalized coupling $g_R$ at scale T
is expressed in terms of the bare (lattice) coupling $g$ as
\begin{equation}
\frac{1}{g^2_R (T)} \simeq \frac{1}{g^2} - \frac{1}{2\pi} \log
N_t
\label{eq:six}
\end{equation}
$N_t$ is the lattice size in the (Euclidean) time direction. In
order to test our lattice approach we should
verify the quasiclassical prediction for
the exponential slope in $1 \over {g^2}$
\begin{equation}
\frac{F_{SP} (T)}{T} \simeq (8 \omega   N_t ) \frac{1}{g^2}
+ {\rm const}
\label{eq:seven}
\end{equation}
if $g^2$ is small enough.  A similar computation for the standard
model would be sufficient to control at least the order of
magnitude of sphaleron transitions close to the electroweak
phase transition and this would improve our knowledge
of this transition rate considerably.
\par
In order to determine the sphaleron free energy we shall
measure the probability distribution of the CS number (after
cooling). The CS number itself is given by
\begin{equation}
N_{CS} = \frac{1}{2\pi} \int dx A_1
\label{eq:eight}
\end{equation}
with the vector potential
\begin{equation}
A_\mu = \partial_\mu \alpha - \sin^2 \theta /2\, \partial_\mu \varphi
\label{eq:nine}
\end{equation}
where $\alpha$ is an arbitrary gauge parameter. Convenient choices
would be $\alpha = 0$ for fields close to the vacuum at  $\theta = \pi$,
 or $\alpha = \varphi$ for fields in the upper hemisphere.
For a study of the sphaleron configurations the latter choice is
appropriate.
\par
In our lattice approach we do not use this expression directly
but determine $A_\mu$
\begin{equation}
A_\mu = \frac{1}{2i} (\chi^+ \partial_\mu \chi -(\partial_\mu
\chi^+) \chi )
\label{eq:ten}
\end{equation}
through the corresponding $CP(1)$ variables
\begin{equation}
\chi = e^{i\alpha}  \left( \begin{array}{c}
\sin \theta/2 e^{-i\varphi} \\
\cos \theta/2
\end{array}  \right)
\label{eq:eleven}
\end{equation}
$A_\mu$ is found as the phase of
\begin{equation}
\chi_1^+ \chi_2 \simeq e^{i \cos^2 \theta/2 \partial_\mu \varphi}
\label{eq:twelve}
\end{equation}
where $\chi_1$ and $\chi_2$ are on neighbouring sites of link
$\mu$.
\par
The classical equations of motion are given by
\begin{equation}
\vec L (x) \equiv \partial^2 \vec n - \omega^2 \vec\delta_3 =
\lambda \vec n
\label{eq:thirteen}
\end{equation}
with
\begin{equation}
\vec\delta_3 = (0,0,1)
\label{eq:fourteen}
\end{equation}
$\lambda$ is a Lagrange multiplier to ensure the constraint.
The basic point to observe is that for any classical solution
$\vec L$ is parallel to $\vec n$. Therefore, the quantity
\begin{equation}
D (x) = \vec L (x) \cdot \vec L (x) - [\vec n (x) \cdot \vec L (x)
]^2 \geq 0
\label{eq:fifteen}
\end{equation}
vanishes for any classical solution (including saddle points), and
is positive otherwise. The integral over $D (x)$ is therefore a
convenient new `action' for defining a cooling procedure which does
not drive away configurations from saddle points.
This procedure can be taken over directly to the SU(2) variables
(Higgs as well as gauge part) of the electroweak standard model.
\par
The model was discretized in the standard way on a $N_t \times
128 $ lattice. This corresponds to a temperature $1/N_t$.
We used a heat-bath algorithm to equilibrate the
lattice for 36000000 sweeps and then a ratio of 8 over-relaxation
updates to one heat-bath. We then investigated 10 blocks
 of 36000000 sweeps with 20000 configurations measured per block.
The averages of each of these 10 block measurements were found to be
consistent with being statistically independent and so were used for
an estimate of the statistical error.  The mass gap was measured
from the propagation in the $x$-direction, namely from
the study of the time-slice averages of the correlation
$n_1(x)n_1(x')+n_2(x)n_2(x')$.

To isolate  configurations which had a sphaleron, we first required
that $n_3(x)$ was above 0.5 on average for a region of $x$ of length
$1/\omega$. For those configurations, we measured  $N_{CS}$ in
a window from $x_{max}-1/(2\omega)$
to $x_{max}+1/(2\omega)$, where $x_{max}$ corresponded to the
maximum value of $n_3(x)$.
  Then configurations with $N_{CS} > 0.45$
and $n_3(x_{max}) > 0.90$ were classified as sphalerons.
In order to check the stability of this procedure,
we cooled
the configurations until the average value per link of $D$ was
reduced to a fixed value. Then we used the above selection criteria
on these cooled configurations. The results presented correspond to a
reduction of $D$ by a factor of approximately 10, although we
found that the sphaleron probability was  insensitive to this
threshold value of $D$ (being the same within errors for smaller $g^2$).
  Our cooling algorithm was to replace
site variables by an admixture of the original value and that
which would minimise the action locally. If this procedure
reduced $D$ (see above) then it was accepted - otherwise $D$ was
explicitly minimised which was computationally more demanding
since it involved next-to-nearest neighbour terms etc.
We  tried
a large number of other choices of sphaleron selection criteria - varying
thresholds, windows, cooling rate, cooling duration etc and found
that the changes amounted to an overall constant shift only.
Thus the overall normalisation of the values for $P_{sph}$ quoted
in the table is not significant, but the dependence on $g^2$
and $N_t$ is.
\par
Results for the sphaleron probability  $P_{sph}$
per configuration  as well
as for the mass gap are given in Table 1. We used $\omega = 0.1$
in lattice units for $N_t = 2$ and then varied $N_t$ keeping
$\omega N_t$ fixed to check that the results were consistent.
Thus it is appropriate to quote the sphaleron probability per
unit spatial length (in units of $\omega$). Values of
$P_{sph}/(128 \omega)$ are plotted versus
 $1 \over {g^2}$  in Figure 1. Here it is seen that the results for
different lattice spacings (ie different $\omega$) are all in agreement
with each other. Thus we have determined the dependence on $g^{-2}$ of the
sphaleron production rate per unit length at $\omega / T =0.2$.
This dependence is indeed  comparable to the quasi-classical
expression which is shown by the continuous line (with arbitrary
normalisation - actually $1.87 g^{-2} \exp(-8 \omega /(g^2 T))$ ).
\par
\begin{table}
\caption{Monte Carlo results}
\vskip5mm
\label{tableone}
\begin{tabular*}{130mm}{@{\extracolsep{\fill}}llll} \hline
 $N_t$ & $g^2$ &  $P_{sph}$ & mass gap     \\ \hline
 $2$ & 0.12 & 0.000338(  23) & 0.108  \\
 $2$ & 0.13 & 0.000770(  46) & 0.110  \\
 $2$ & 0.14 & 0.001560(  85) & 0.111  \\
 $2$ & 0.15 & 0.002945(  70) & 0.113  \\
 $2$ & 0.16 & 0.004790( 173) & 0.116  \\
 $2$ & 0.20 & 0.018600( 780) & 0.128  \\
 $2$ & 0.24 & 0.035100(1036) & 0.145  \\
 $2$ & 0.25 & 0.041700(1121) & 0.150  \\
 $3$ & 0.13 & 0.000580(  78) & 0.073  \\
 $3$ & 0.14 & 0.001140( 105) & 0.074  \\
 $3$ & 0.15 & 0.002010( 112) & 0.075  \\
 $4$ & 0.13 & 0.000518(  60) & 0.055  \\
 $4$ & 0.14 & 0.000800(  70) & 0.056  \\
 $4$ & 0.15 & 0.001575( 125) & 0.057  \\
 $4$ & 0.16 & 0.003062( 187) & 0.058  \\      \hline
\end{tabular*}
\end{table}
\par
The quasi-classical calculation of fluctuations around the classical
spaleron gives a result for the spaleron probability~\cite{mottola}
which is expected to be valid if $g^2 T < \omega$. Our results
at  $g^2 < 0.2$ are thus expected to be approximately reproduced
by this approach. As shown in the figure, this is indeed the case.

This close agreement seems a little fortuitous.
One reason is that
a perturbative expansion in terms of lattice parameters does not
work very well in the present parameter range, but only at much
smaller coupling. Thus the measured mass gap and the renormalized coupling
should be used instead of $\omega$ and $g^2$ in eq.(8). So far,
we have not measured the renormalized coupling and cannot make
an estimate of the size of this effect.
A somewhat smaller value for the sphaleron
free energy might also be expected because of the finite range
of attraction of the sphaleron in our cooling procedure.
However, we find very consistent results as the extent of the smoothing
is varied over a wide range. This suggests that such an effect should
be small.
\par
In conclusion, we have demonstrated a method for measuring
Chern--Simons transitions which can be carried over to study
the bosonic sector of the electroweak standard model. This will
allow checking various estimates based on resummed perturbation
theory for effective potentials. The outcome is crucial for
judging the viability of approaches for generating the baryon
number of the universe at the electroweak phase transition.
\vskip5mm
\par
\vskip1.0cm \noindent

\newpage
\begin{figure}
\centering
\vspace{15cm}
\includegraphics{jkcm_f.ps}
\caption{
The probalitity of a sphaleron-like configuration per unit spatial
length as a function of $1/g^2$.
}
\end{figure}
\end{document}